\documentstyle[12pt]{article}

\begin{document}

\newcommand{\Lslash}[1]{ \parbox[b]{1em}{$#1$} \hspace{-0.8em}
                         \parbox[b]{0.8em}{ \raisebox{0.2ex}{$/$} }    }
\newcommand{\Sslash}[1]{ \parbox[b]{0.6em}{$#1$} \hspace{-0.55em}
                         \parbox[b]{0.55em}{ \raisebox{-0.2ex}{$/$} }    }
\newcommand{\Mbf}[1]{ \parbox[b]{1em}{\boldmath $#1$} }
\newcommand{\mbf}[1]{ \parbox[b]{0.6em}{\boldmath $#1$} }
\newcommand{\beq}{\begin{equation}}
\newcommand{\eeq}{\end{equation}}
\newcommand{\beqa}{\begin{eqnarray}}
\newcommand{\eeqa}{\end{eqnarray}}
\newcommand{\skipfields}{\!\!\!\!\! & \!\!\!\!\! &}

\newcommand{\gsim}{\buildrel > \over {_\sim}}
\newcommand{\lsim}{\buildrel < \over {_\sim}}
\newcommand{\ie}{{\it i.e.}}
\newcommand{\eg}{{\it e.g.}}
\newcommand{\cf}{{\it cf.}}
\newcommand{\etal}{{\it et al.}}
\newcommand{\gev}{{\rm GeV}}
\newcommand{\jpsi}{J/\psi}
\newcommand{\order}[1]{${\cal O}(#1)$}
\newcommand{\eq}[1]{(\ref{#1})}

\newcommand{\FigDiagram}{Fig.\ 1}
\newcommand{\FigSpline}{Fig.\ 2}
\newcommand{\FigRegions}{Fig.\ 3}
\newcommand{\FigAdependence}{Fig.\ 4}
\newcommand{\FigTrapez}{Fig.\ 5}
\newcommand{\FigRescaling}{Fig.\ 6}
\newcommand{\FigInverse}{Fig.\ 7}

\newcommand{\ptr}{p_T}
\newcommand{\as}{\alpha_s}
\newcommand{\xbj}{x_{\rm Bj}}

\newcommand{\PL}[3]{Phys.\ Lett.\ {#1} ({#3}) {#2}}
\newcommand{\NP}[3]{Nucl.\ Phys.\ {#1} ({#3}) {#2}}
\newcommand{\PRD}[3]{Phys.\ Rev.\ D {#1} ({#3}) {#2}}
\newcommand{\PRL}[3]{Phys.\ Rev.\ Lett.\ {#1} ({#3}) {#2}}
\newcommand{\ZPC}[3]{Z. Phys.\ C {#1} ({#3}) {#2}}
\newcommand{\PRe}[3]{Phys.\ Rep.\ {#1} ({#3}) {#2}}

\newcommand{\mobi}{mobility distribution}

%
%***** For Document style: article
%\pagestyle{headings}
\newcommand{\LevelOne}[1]{ \newpage \subsection*{#1} }
\newcommand{\LevelTwo}[1]{ \subsubsection*{#1} }
\newcommand{\LevelThree}[1]{ \paragraph*{#1} }
\newcommand{\AppendixLevel}[2]{
   \newpage
   \subsection*{Appendix #1 -- #2}
   \markright{Appendix #1}
   \addcontentsline{toc}{subsection}{\protect\numberline{#1}{#2}} }

\begin{titlepage}
\begin{flushright}
        NORDITA-96/20 P
\end{flushright}
\begin{flushright}
        hep-ph/9604305 \\ \today
\end{flushright}
\vskip .8cm
\begin{center}
{\Large Nuclear Dependence of Structure Functions in Coordinate Space}
\vskip .8cm
 {\bf P. Hoyer and M. V\"anttinen}
\vskip .5cm
NORDITA\\
Blegdamsvej 17, DK-2100 Copenhagen \O
\vskip 1.8cm
\end{center}

\begin{abstract} \noindent
The momentum distributions of partons in bound nucleons are known
to depend significantly on the size of the nucleus.
The Fourier transform of the momentum ($\xbj$)
distribution measures the
overlap between Fock components of the nucleon wave function which differ by
a displacement of one parton along the light cone.
The magnitude of the overlap thus determines the average range
of mobility of the parton in the nucleon. By comparing the
Fourier transforms of structure functions for several nuclei we study the
dependence of quark mobility on nuclear size. We find a surprisingly small
nuclear dependence ($<2\%$ for He, C and
Ca) for displacements $t=z \lsim 2.5$
fm, after which a nuclear suppression due to shadowing sets in.
The nuclear effects observed in momentum space for
\mbox{$\xbj \lsim 0.4$}
can be understood as a reflection of only the large distance
shadowing in coordinate space.
\end{abstract}

\end{titlepage}

\setlength{\baselineskip}{7mm}

%111111111111111111111111111111111111111111111111111111111111111111111111111
\subsection*{1. Introduction}

The difference between the
quark structure functions of nuclei and those of free nucleons,
first observed in 1982 by the EMC collaboration \cite{EMC82}, has generated
considerable experimental and theoretical interest \cite{Arneodo}. %PH
The measured
nuclear structure function gives direct information about how the momentum
distribution of quarks in nucleons is modified by nuclear binding effects.
Several models have been proposed for such modifications, many of them %PH
based on a picture of the nuclear wave function in coordinate space, from
which modifications of the momentum space distribution are surmised.

The inclusive lepton scattering measurements determine the
single parton distributions, but do not constrain parton-parton correlations
in a model-independent way. Hence the inclusive data can only partially
constrain theoretical models for the nuclear effect. %PH

Here we wish to 
study the nuclear effects in coordinate space. %PH
The relation between quark
distributions in momentum and coordinate space has been
known for a long time, and involves no further model dependence than is
needed for the usual QCD interpretation of the experimental structure
functions 
\cite{CS,BB,Braun}. %PH
The phenomenological discussions of 
nuclear effects on %PH
parton distributions have
nevertheless concentrated almost uniquely on momentum space (see, however,
Ref.\ \cite{Llewellyn}). %PH
While the momentum and coordinate space
descriptions are in principle equivalent, insight into the physical 
mechanisms %PH
may benefit from viewing the quark distributions
in both spaces. For example, the standard explanation of the small $x$
shadowing effect as due to the soft scattering on the nucleus of $q \bar q$
pairs created by the virtual photon well upstream of the target
is most naturally discussed in coordinate space \cite{DBH}. %PH

The cross section of  deep inelastic lepton
scattering (DIS) as a function of the photon virtuality $-Q^2$
and the Bjorken variable
$\xbj \equiv x = Q^2/2m\nu$, %MV
where $m$ is the
nucleon mass and $\nu$ is the energy of the photon in the
target rest frame, can be parametrized as
\beq
  \frac{d\sigma}{dxdQ^2} 
%  & = &
  = \frac{4\pi\alpha^2}{Q^4} \frac{F_2(x,Q^2)}{x}
%        \nonumber \\
%  &   & \times
        \left[
        1 - y - \frac{xym}{2E}
        + \frac{y^2}{2} \frac{1 + 4m^2x^2/Q^2}{1 + R(x,Q^2)}
        \right] ,
\eeq
in terms of the structure function $F_2(x,Q^2)$
and the ratio $R(x,Q^2) = \sigma_L/\sigma_T$ of the cross
sections induced by longitudinally and transversely polarized
virtual photons
(in this formula, $y=\nu/E$ is the fraction of initial lepton energy carried
by the photon). Measurements on a variety  of nuclear targets
$A$ have shown that the  $A$-dependence of $R(x,Q^2)$ is
weak
\cite{Arneodo}, %PH
so that the ratio of $F_2$ structure functions for
different targets $A_1, A_2$ is approximately given by the
ratio of measured cross sections:
\mbox{$ F_2^{A_1}/F_2^{A_2} = d\sigma^{A_1}/d\sigma^{A_2} $}.
The $Q^2$ dependence of \mbox{$ F_2^{A_1}/F_2^{A_2}$}
is also known to be weak 
\cite{Arneodo} %PH
and will not be discussed below.

According to perturbative QCD at lowest order in $\as$, the
$F_2$ structure function is given by
\beq
  F_2(x,Q^2)=\sum_i e_i^2 x [q_i(x,Q^2)+\bar q_i(x,Q^2)] \label{f2}
\eeq
where $q_i(x,Q^2)$ is a quark distribution in momentum
space, \ie\ the probability that a quark of flavor $i$
(having electric charge $e_i$ in units of $e$) carries
a %PH
light-cone fraction $x$ of the nucleon momentum.

The deep inelastic $eN \to e'X$ cross section is related to the forward
$\gamma^*N \to \gamma^*N$ hadronic matrix element
\beq
  T_{\mu\nu} = \int d^4y \exp(iq \cdot y)
  \langle P| T \, [j_\mu(y)^\dagger j_\nu(0)] |P\rangle 
  \label{tmunu}
\eeq
through
\beqa
  {\rm Im} \, T_{\mu\nu}
  & = & 4\pi^2 \left[
        -F_1(x,Q^2) \left( g_{\mu\nu} - \frac{q_\mu q_\nu}{q^2} \right) %MV
        \right. \nonumber \\
  &   & \left. \mbox{} + F_2(x,Q^2) \frac{1}{p\cdot q} %MV
        \left( p_\mu - \frac{p\cdot q}{q^2}q_\mu \right)
        \left( p_\nu - \frac{p\cdot q}{q^2}q_\nu \right)
        \right] ,
\eeqa
where $F_1 = F_2/2x$ at lowest order in $\alpha_s$.
In the frame where the virtual photon momentum is
$q=(\nu,\vec 0_\perp,\sqrt{\nu^2+Q^2})$,
its light-cone components \mbox{$q^\pm \equiv q^0\pm q^3$} are
\beq
  q^+ \simeq 2\nu, \;\;
  q^- \simeq -\frac{Q^2}{2\nu} = -mx.
\eeq
The Fourier transform in \eq{tmunu} then
implies a resolution in coordinate space of $\delta y^-\simeq
1/2\nu$, $\delta y^+\simeq 1/mx$. In the scaling limit $Q^2,\
\nu \to \infty$ with $x$ fixed, the most relevant separations $y$ between the
photon currents in \eq{tmunu} are light-like distances of order the
`Ioffe length' $1/2mx$ \cite{Ioffe}.

A visualization of ${\rm Im}\, T_{\mu\nu}$ in coordinate space
is given in \FigDiagram. In the target rest frame, where
$P=(m,\vec 0)$, the target is moving along the $y^0$ axis, while
the photon enters along the positive light cone. 
The imaginary part of $T_{\mu\nu}$ measures the overlap of
two Fock states of the target, which have identical parton content
except for the quark struck by the photon, whose $y^+$ coordinate
is offset by $\delta y^+ \simeq 1/mx$. The magnitude of
the overlap is a measure of the mobility of
the struck quark in the target wave function.

The precise meaning of the Fourier transform of the quark structure function
is given by the operator product expansion through the relation
\cite{CS,BB,Braun} %PH
\beqa
  \lefteqn{\langle P|\bar\Psi(y^+)
  \gamma^- %PH
  \Psi(0)|P\rangle_{\mu^2}} 
  \nonumber \\
  & = &
  2m\int_0^A dx \left[ q(x,\mu^2)
  \exp \left( \frac{imxy^+}{2} \right)
  - \bar q(x,\mu^2) \exp \left( -\frac{imxy^+}{2} \right) \right] ,
  \label{ope}
\eeqa
which here is formulated in $A^-=0$ gauge at a renormalization scale $\mu^2$
and in the rest frame of the target. The left-hand side
of \eq{ope} measures (\cf\ \FigDiagram)
the interference of target Fock states after either a quark is
displaced along the light cone from 0 to $y^+=y^0 + y^3$, or an
antiquark is displaced the opposite distance. The
relative minus sign is due
to Fermi statistics. The kinematic upper limit of the $x$-integral on the
right-hand side
of \eq{ope} is the atomic number $A$ of the target, due to the scale
$m=m_N$ used in the definition of $x$.

Experimental information on $A$-dependence is available mainly for the $F_2$
structure function. Subtracting from \eq{ope} the same relation with
\mbox{$y^+ \to -y^+$} we have
\beqa
  \lefteqn{\langle P|\bar\Psi(y^+)
  \gamma^- %PH
  \Psi(0)|P\rangle_{\mu^2}
  - (y^+\to -y^+)}
  \nonumber \\
  & = & 4im\int_0^A dx[q(x,\mu^2) + \bar q(x,\mu^2)]
        \sin \left( \frac{mxy^+}{2} \right) .
  \label{f2ft}
\eeqa
After summing over the quark flavors weighted by $e_i^2$, the integral
can be evaluated using the measured $F_2$ structure function for
a range of targets.

\subsection*{2. Numerical Analysis} %PH

The nuclear target effects on $F_2$ can be expressed through the ratio
\beq
  R_A(x)= \frac{F_2^A(x,Q^2)}{(A/2) F_2^D(x,Q^2)} \label{ra}
\eeq
of the structure function measured on a nuclear target $A$ to that on the
deuteron D. Experiments show that the ratio (\ref{ra}) is practically
independent of $Q^2$. There is data on $R_A$ for $A=$ He, C and
Ca from both CERN \cite{NMC} and SLAC \cite{E139}, which together cover
the ranges \mbox{$0.0035 < x < 0.88$}, \linebreak
\mbox{$0.00015 < x < 0.8$} and
\mbox{$0.0035 < x < 0.8$}, respectively.
Measurements on heavier nuclei have been done for different choices
of $A$ in different experiments, and will not be used here. The available
data on $R_C$ is shown in \FigSpline. In our evaluations of the Fourier
transform (\ref{f2ft}) we fitted the data with a smooth curve (solid
line). We also used an integration algorithm based on the discrete data
points to obtain an error estimate.

We used the fit of the $F_2^D$ structure function given by
the NMC Collaboration \cite{NMC:F2D}, evaluated at $Q^2=5\ \gev^2$
(a typical value in experimental determinations of $R_{A/D}(x)$).
This fit also includes SLAC and BCDMS data.
%PH
The $F_2^A(x,Q^2=5\ \gev^2)$ structure functions are obtained as
a product of the NMC fit for $F_2^D$ and our fit of the ratio (\ref{ra}). The
Fourier transform (\ref{f2ft}) then gives the quark
`mobility' distribution in coordinate space for various nuclei,
\beq
  Q^A(y^+,Q^2) \equiv \int_0^1 dx F_2^D(x,Q^2)
  R^A(x)\frac{\sin(mxy^+/2)}{x} . \label{qa}
\eeq
Using the quark \mobi s $Q^A(y^+,Q^2)$ we can then form the target
ratio in coordinate space,
\beq
  R_A(y^+,Q^2)= \frac{Q^A(y^+,Q^2)}{(A/2) Q^D(y^+,Q^2)} .  \label{raco}
\eeq

Strictly speaking, the upper limit of the $x$-integral in \eq{qa} should
be at $x=A$. The large $x$ region is, however, unimportant in the Fourier
transform due to the small size of $F_2^A$ in this region. This is
illustrated in \FigRegions, which shows the contribution
to the integral in
\eq{qa} from various regions of $x$. The fact that the large $x$ region is
insignificant also implies that the effects of nuclear Fermi motion are
suppressed in coordinate space (at moderate values of $y^+$).

From \FigRegions\ one can already anticipate important cancellations
of the $A$-dependence in the Fourier integral.
For $y^+ = y^0 + y^3$ below 2 fm,
the `anti-shadowing' nuclear enhancement region beats against the EMC
suppression region, while for $y^+$ around 5 fm anti-shadowing tends
to offset the shadowing suppression at low $x$. At large light-cone
distances the low $x$ region dominates the integral, and the shadowing effect
will prevail.

Assuming a power law dependence
$F_2^A(x) \simeq c_A x^{-\beta}$ for small $x$ gives
% in fact %PH
\beq
  Q^A(y^+\to \infty)
  \simeq c_A \left( \frac{m y^+}{2} \right)^{\beta}
  \int_0^\infty du \, \frac{\sin u}{u^{1+\beta}} .
  \label{highz}
\eeq
If the power $\beta$ is independent of $A$, \ie, if
$R_A(x)$ in \eq{ra} approaches 
the constant %PH
$2c_A/Ac_D$ for $x \to 0$, the
coordinate space ratio $R_A(y^+)$ in \eq{raco} will approach
the same constant %PH
$2c_A/Ac_D$ for $y^+ \to \infty$. This limiting value is numerically
reached only for very large $y^+$, however
(see \FigAdependence\ below). 

In our numerical calculation, we have set
$F_2(x,Q^2) = F_2(x=0.01,Q^2)$ for $x < 0.01$,
motivated by the $x$ range of the NMC parametrization for $F_2^D$.
We have assumed the ratio $R_A(x)$ to approach a constant for $x \to 0$,
and verified that the results presented here are insensitive to
the precise value of $R_A(0)$ and to the behaviour of $F_2^D$ at
$x < 0.01$, where HERA data \cite{HERA} actually show a rapid rise of
$F_2^D$ with decreasing $x$.

\subsection*{3. Results and Discussion} %PH

The coordinate space target ratio $R_A(y^+,Q^2=5\;\gev^2)$ %PH
(\ref{raco}) is shown in \FigAdependence\ %PH
for $A=$ He, C and Ca. 
Even for the 
heaviest %PH 
(Ca) nucleus,
the ratio %PH 
is within 2 \% of unity for $y^+ = y^0 + y^3 \lsim 5$ fm. At larger distances,
the shadowing effect is clearly visible.

\FigTrapez\ shows error estimates for
$R_C(y^+, Q^2=5\;\gev^2)$ %PH
derived from the actual data points
and errors of the measured momentum space ratios $R_C(x)$
by using trapezoidal rule integration \cite{TRAPER}.
For large $y^+$, the finite spacing of the data
points in $x$ makes any error estimate dependent on how smooth $R_C(x)$ is
assumed to be. 
%The Fourier transform is not, however, sensitive to %PH
%reasonable (smooth) extrapolations of the data to $x=0$ such as those %PH
%shown by the dashed curves in Fig.\ \ref{RC}. %PH
The result for $R_C(y^+,Q^2)$
shown in \FigAdependence\ should 
thus %PH
be reliable at
least for $y^+ \lsim 7$ fm.
We have also verified that our results are insensitive to the
value of $Q^2$ used in evaluating the structure function of the deuteron.

At low $y^+$, the quark mobility 
distribution %PH 
$Q^A(y^+)$ is readily seen
from \eq{qa} to be proportional to $y^+$,
\beq
  Q^A(y^+\to 0) \simeq y^+ \frac{m}{2} \int_0^1 dx F_2^A(x) ,
\label{lowz}
\eeq
where the 
integral %PH 
measures the total fraction of target momentum
carried by quarks. Previous careful estimates of the $A$-dependence of
this fraction \cite{Arneodo}, which took into account
finite energy effects in the
data, gave results compatible with no nuclear dependence,
\beq
  \int_{0.0035}^{0.80} dx [F_2^{\rm Ca}-F_2^{D}]
  = (-1.5 \pm 0.4 \pm 1.4) \cdot 10^{-3} ,
  \label{f2int}
\eeq
to be normalized by \mbox{$\int dx F_2^D \simeq 0.15$}.
%PHb
%Hence, taking into account higher twist effects, 
%we must conclude that the
%$R_A(y^+)$ ratios in \FigAdependence\ actually are compatible
%with unity in the whole region \mbox{$y^+ \lsim 5$ fm.}
As seen from \FigAdependence, the deviation of
$R_{\rm Ca}(y^+)$ from unity is no larger than it is at $y^+=0$ in the whole
region \mbox{$y^+\lsim 5$ fm.}
Such a weak $A$-dependence of the structure
function in coordinate space appears rather surprising,
given %PH 
that the EMC
effect is of \order{10 \ldots 15\%} for C and Ca nuclei in momentum
space. As discussed above, the weakness of the 
%PHb
nuclear effect in coordinate space is due to cancellations between 
nuclear enhancement and suppression regions in momentum space.
%PHe

%Some of the models discussed in the context of the EMC effect
%\cite{swelling} postulated
%that the effective nucleon radius would be larger by up to 30\% in a
%nucleus (`nucleon swelling'). The Fourier 
%transform (\ref{qa}) %PH
%provides 
%the exact relation between distributions %PH 
%in coordinate and
%momentum space. We find that a simple rescaling of light-cone
%distances, $Q^A(y^+) \propto Q^D(\xi y^+)$, leads to an incorrect
%shape of the ratio $R_A(y^+)$ for $y^+\lsim 3$ fm, as shown in
%\FigRescaling, unless $1-\xi \lsim 2\%$. Such a model in
%any case fails to explain the shadowing region of large $y^+$.

Some of the models discussed in the context of nuclear effects
postulated that the effective radius of a bound
nucleon would be larger by up to 30\% compared to that of a free nucleon
(``nucleon swelling'' \cite{swelling}).
The Fourier transform (\ref{qa}) provides the exact relation
between distributions in coordinate and momentum space.
We find that a simple assumption $Q^A(y^+) \propto Q^D(\xi y^+)$
leads to an incorrect shape of the ratio $R_A(y^+)$
for $y^+\lsim 3$ fm, as shown in \FigRescaling, unless
$1-\xi \lsim 2\%$. Such a model in any case fails to describe
the shadowing region of large $y^+$.

This analysis is not directly applicable to
the ``$Q^2$ rescaling'' models \cite{rescaling}, where a
nuclear effect arises because the effective
value of $Q^2$ is taken to be different for bound and free nucleons
due to their different radii.

\FigAdependence\ suggests that ignoring a
possible $\lsim 2\%$ effect for $y^+ \lsim 5$ fm, the main nuclear effect in
coordinate space is the shadowing phenomenon at large $y^+$. The physical
reason for shadowing in DIS is well understood at a qualitative level,
and quantitative models have been
successfully constructed \cite{PH:Paris}.
It may thus be of some interest to see how much of
the structure in momentum space can be ascribed solely to shadowing
in coordinate space. %PH
This can be studied be evaluating the 
transform inverse to the one in \eq{qa}, %PH
\beq
  \widetilde F_2^A(x)= \frac{m}{\pi}\; x %PH
  \int_0^\infty dy^+
  \widetilde Q^A(y^+,w) %PH
  \sin(mxy^+/2) \label{invqa}
\eeq
using a function 
$\widetilde Q^{A}(y^+,w)$ %PH
defined as
\beq
 \widetilde Q^{A}(y^+,w) = \left\{ \begin{array}{ll} %PH
              Q^D(y^+), & y^+ < w, \\
              Q^A(y^+), & y^+ \ge w,
              \end{array} \right.  \label{qmod}
\eeq
where $w = {\cal O}$(5 fm). 
%PHb
Substituting the definition (\ref{qa}) 
of $Q^{D,A}(y^+)$ and interchanging the order of integration one
obtains an expression
suitable %MV
for numerical evaluation,
\beqa
  \widetilde F_2^A(x,w)
  & = & F_2^A(x)+\frac{x}{\pi} \int_0^1 dx' \left[Q^A(x')-Q^D(x') \right] 
  \nonumber \\
  &   & \times \left\{ \frac{\sin[m(x'+x)w/2]}{x'+x} -
                      \frac{\sin[m(x'-x)w/2]}{x'-x} \right\}. \label{invnum}
\eeqa
The corresponding momentum-space target ratio
\beq
  \widetilde R_A(x,w) = \frac{\widetilde F_2^A(x,w)}{(A/2) F_2^D(x)}
  \label{tilra}
\eeq
for two choices of $w$ is compared with $R_A(x) = \widetilde R_A(x,w=0)$
in \FigInverse. It can be seen that the nuclear effects
observed in the data for $x \lsim 0.4$ (which includes the anti-shadowing
enhancement and the beginning of the EMC suppression) can be obtained by
assuming no other nuclear effect in coordinate space than the shadowing
for $y^+ \gsim 5$ fm. The nuclear effects at larger values of $x$ do,
however, depend sensitively also on the small ($\lsim 2 \%$) effects
suggested by the data at light-cone distances $y^+ \lsim 5$ fm.

\subsection*{4. Summary} 

We have studied the nuclear effects on parton distributions in coordinate
space by Fourier transforming the measured momentum space ($\xbj$)
distributions. Parton distributions in coordinate space can be rigorously
defined using the Operator Product Expansion. Intuitively, they measure
parton mobility in the target wave function, in terms of an average
overlap between wave function components where one parton has been offset
the given distance along the light cone.

The parton distributions at large light-cone distances $y^+$ are sensitive
only to the momentum distribution at small $\xbj$, and thus reflect the
well-known shadowing phenomenon. For $y^+ \lsim 5$ fm
(\ie, for $y^0 = y^3 \lsim 2.5$ fm), on the other hand,
we found the nuclear effects to be surprisingly weak ($<2 \%$ for $A$ =
He, C and Ca). Numerically, this is due to cancellations in the Fourier
integral between regions of nuclear enhancement (anti-shadowing) and
suppression (shadowing, EMC effect).
Conversely, the observed nuclear dependence in momentum space
reflects the mixing of effects from long and short distances.
If the small nuclear effect at
moderate light-cone distances is interpreted as an effective `swelling' of
the nucleons in the nucleus, only a $\lsim 2\%$ increase in the nucleon
radius can be accomodated.

Through an inverse transform we verified that the nuclear effects observed
for $\xbj \lsim 0.4$ can be obtained solely from the shadowing
phenomenon at \mbox{$y^+ \gsim 5$ fm} in coordinate space. The effects seen at
larger $\xbj$ depend sensitively also on weak nuclear dependencies at
\mbox{$y^+ \lsim 5$ fm}.

\subsection*{Acknowledgements}

We would like to thank V. Braun and H. Pirner for helpful
discussions.
%PHe

%111111111111111111111111111111111111111111111111111111111111111111111111111

\pagebreak

\begin{center} {\Large FIGURE CAPTIONS} \end{center}

\bigskip

%MV Figure captions modified on 3 April

\begin{description}

\item[Fig. 1.]
A space-time picture of 
the imaginary part of the matrix element $T_{\mu\nu}$ %PH
of \eq{tmunu} %MV
for forward virtual %PH
photon-nucleon scattering.

\item[Fig. 2.]
Our fit of the structure function ratio $R_C(x)$ for carbon %PH
compared with data. Squares: NMC data \cite{NMC};
diamonds: SLAC-E139 data \cite{E139}.
The solid curve is the result of a least-squares cubic spline fit
with interior knots at $x = 0.002$, 0.02, 0.2 and 0.75
and an extra ``data point'' $R_C(x=0)=0.85$ added at $x=0$,
with error $\Delta R_C(x=0) = 0.01$. The 
dashed %PH 
curves show
the effect of varying $R_C(x=0)$ between 0.7 and 0.95.
(a) Logarithmic horizontal scale, %PH
(b) linear horizontal scale. %PH

\item[Fig. 3.]
The solid curve shows the \mobi\ $Q^D(y^+,Q^2=5\;\gev^2)$
for deuterium. The dotted curves show the contributions to $Q^D(y^+,Q^2)$
from different regions in the $x$ integral of 
\eq{qa}.

\item[Fig. 4.]
The coordinate space %PH
ratio $R_A(y^+,Q^2=5\;\gev^2)$ for $A=$ He, C and Ca as
obtained by using cubic-spline fits of the $R_A(x)$ data
(\cf\ \FigSpline\ for carbon) and the NMC parametrization
of $F_2^D(x,Q^2)$ \protect\cite{NMC:F2D}.
At very large $y^+$, $R_A(y^+)$ tends to the constant value
$R_A(x=0)$, which has here been set to 0.92, 0.85 and 0.75
for He, C and Ca, respectively.

\item[Fig. 5.]
An estimated error band for  $R_C(y^+,Q^2=5\;\gev^2)$.
The solid line is as in \FigAdependence.
The dotted lines show error estimates
based on the experimental errors of $R_C(x)$
and trapezoidal rule integration \protect\cite{TRAPER}.
At \mbox{$y^+ \gsim$ 7 fm}, the
trapezoidal rule integration becomes unreliable because of the finite
spacing of the data points
in $x$. %PH

\item[Fig. 6.]
Solid curve: $R_C(y^+,Q^2=5\;\gev^2)$ 
from \FigAdependence. %PH
%as obtained by using a spline fit %PH
%of the $R_C(x)$ data. The fit is as in \FigSpline. %PH
Dashed %PH 
curves: $R_C(y^+)$ as obtained by setting
$Q_C(y^+) = Q_D(\xi y^+)$.

\item[Fig. 7.]
The momentum space ratio $\widetilde R_{C}(x,w,Q^2=5\;\gev^2)$
obtained from the
inverse transform (\ref{invnum}) using the modified coordinate space
distribution
$\widetilde Q_{A}(y^+,w)$ (\ref{qmod}), in which all nuclear
effects are eliminated for \mbox{$y^+ < w$}.

\end{description}


\begin{thebibliography}{99}

\bibitem{EMC82}
J. J. Aubert \etal, EMC Coll.: \PL{105B}{322}{1982}.

\bibitem{Arneodo}
M. Arneodo: \PRe{240}{301}{1994}.

\bibitem{CS} %PH
J. C. Collins, D. E. Soper: \NP{B194}{445}{1982}.

\bibitem{BB} %PH
I. I. Balitsky, V. M. Braun: \NP{B311}{541}{1989}.

\bibitem{Braun}
V. Braun, P. G\'ornicki, L. Mankiewicz: \PRD{51}{6036}{1995}.

\bibitem{Llewellyn}
C. H. Llewellyn Smith: \NP{A434}{35c}{1985}.

\bibitem{DBH}
V. Del Duca, S. J. Brodsky, P. Hoyer: \PRD{46}{931}{1992}.

\bibitem{Ioffe}
B. L. Ioffe: \PL{30B}{123}{1969}.

\bibitem{NMC}
P. Amaudruz \etal, NMC Coll.: \NP{B441}{3}{1995};
M. Arneodo \etal, NMC Coll.: \NP{B441}{12}{1995}.

\bibitem{E139}
J. Gomez \etal, SLAC-E139 Coll.: \PRD{49}{4348}{1994}.

\bibitem{NMC:F2D}
P. Amaudruz \etal, NMC Coll.: \PL{B295}{159}{1992}.

\bibitem{HERA}
M. Derrick \etal, ZEUS Coll.: \ZPC{65}{379}{1995};
T. Ahmed \etal, H1 Coll.: \NP{B439}{471}{1995}.

\bibitem{TRAPER} We used the TRAPER subroutine from the CERN 
MATHLIB library.

%\bibitem{swelling}
%S. Fredriksson: \PRL{52}{724}{1984};
%
%A. W. Hendry, D. B. Lichtenberg, E. Predazzi:
%\PL{B136}{433}{1984}; Nuovo Cim.\ {\bf 92A}, 427 (1986);
%
%V. Barone, E. Predazzi: Nuovo Cim.\ {\bf 99A}, 661 (1988);
%
%L. L. Frankfurt, M. I. Strikman:
%Sov.\ J. Nucl.\ Phys.\ {\bf 41}, 1585 (1985); \PRe{160}{235}{1988};
%
%V. Barone \etal: \ZPC{58}{541}{1993}.

\bibitem{swelling} J. V. Noble, \PRL{46}{412}{1981};
A. W. Hendry, D. B. Lichtenberg and E. Predazzi, \PL{136B}{433}{1984}.

\bibitem{rescaling} O. Nachtmann and H. J. Pirner, \ZPC{21}{277}{1984};
R. L. Jaffe, F. E. Close, R. G. Roberts and G. G. Ross, \PL{134B}{449}{1984}.

\bibitem{PH:Paris}
For a recent review, see
P. Hoyer: {\em Interactions on Nuclei},
in Proc.\ of the Workshop on Deep Inelastic Scattering and QCD,
Paris, 24-28 April 1995, eds.\ J. F. Laporte and Y. Sirois, p.\ 127.

\end{thebibliography}
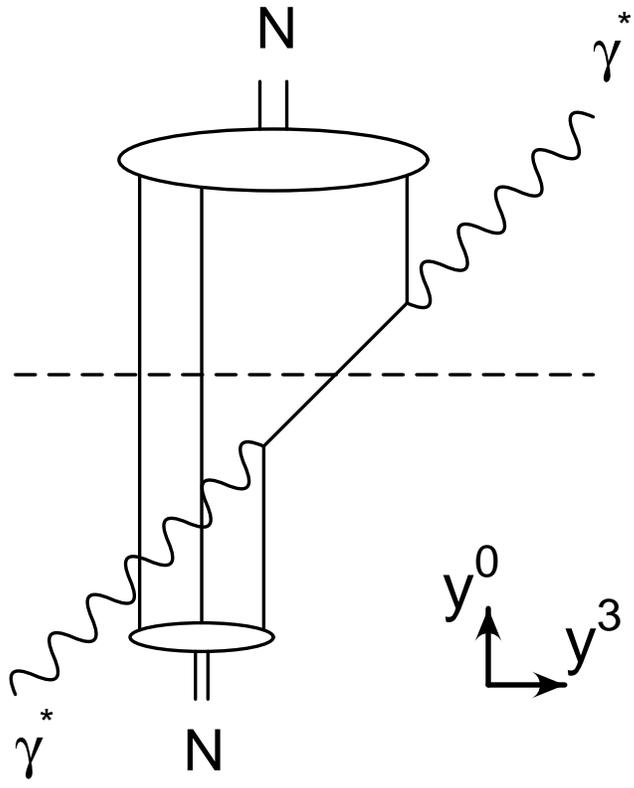
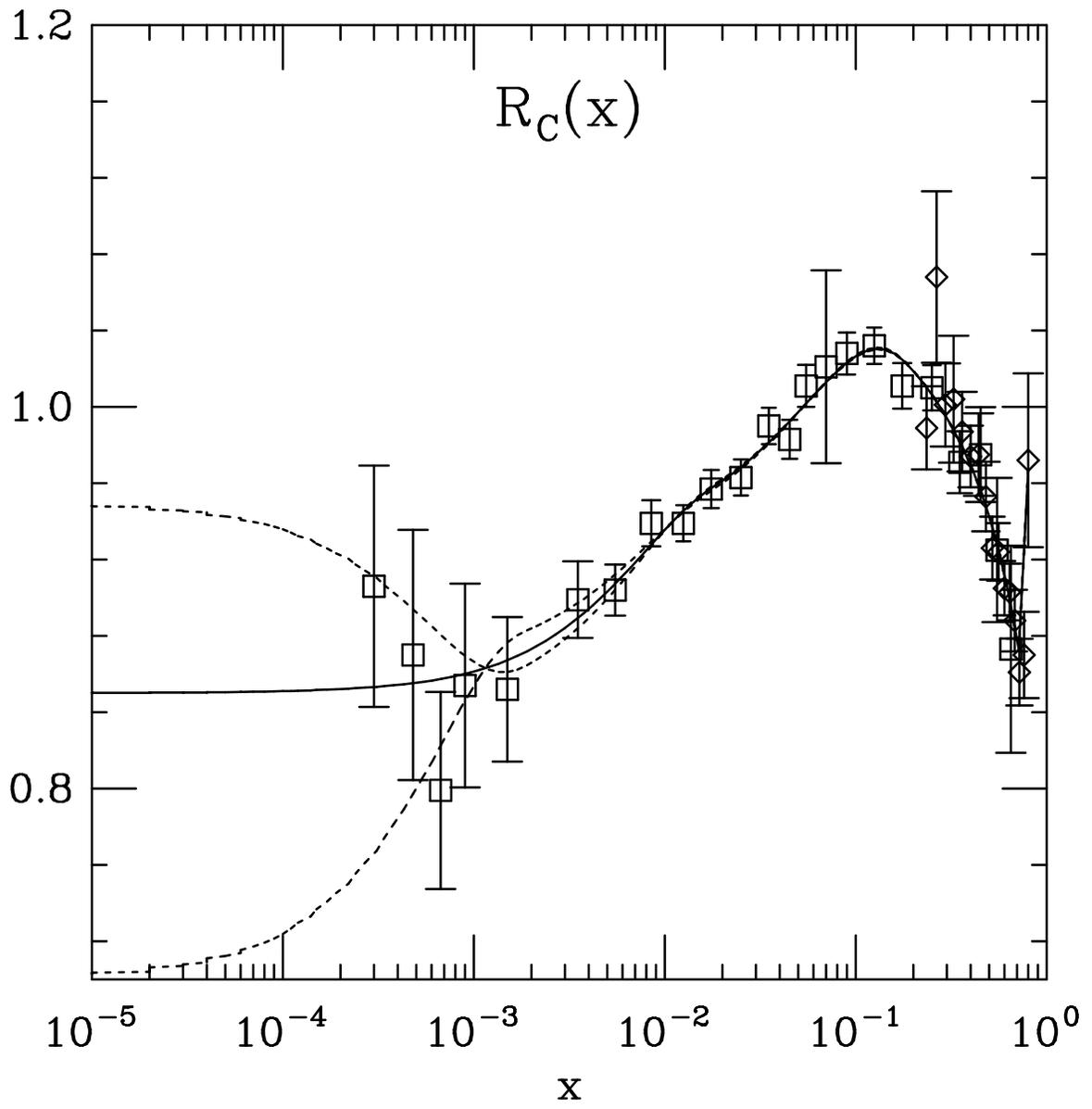
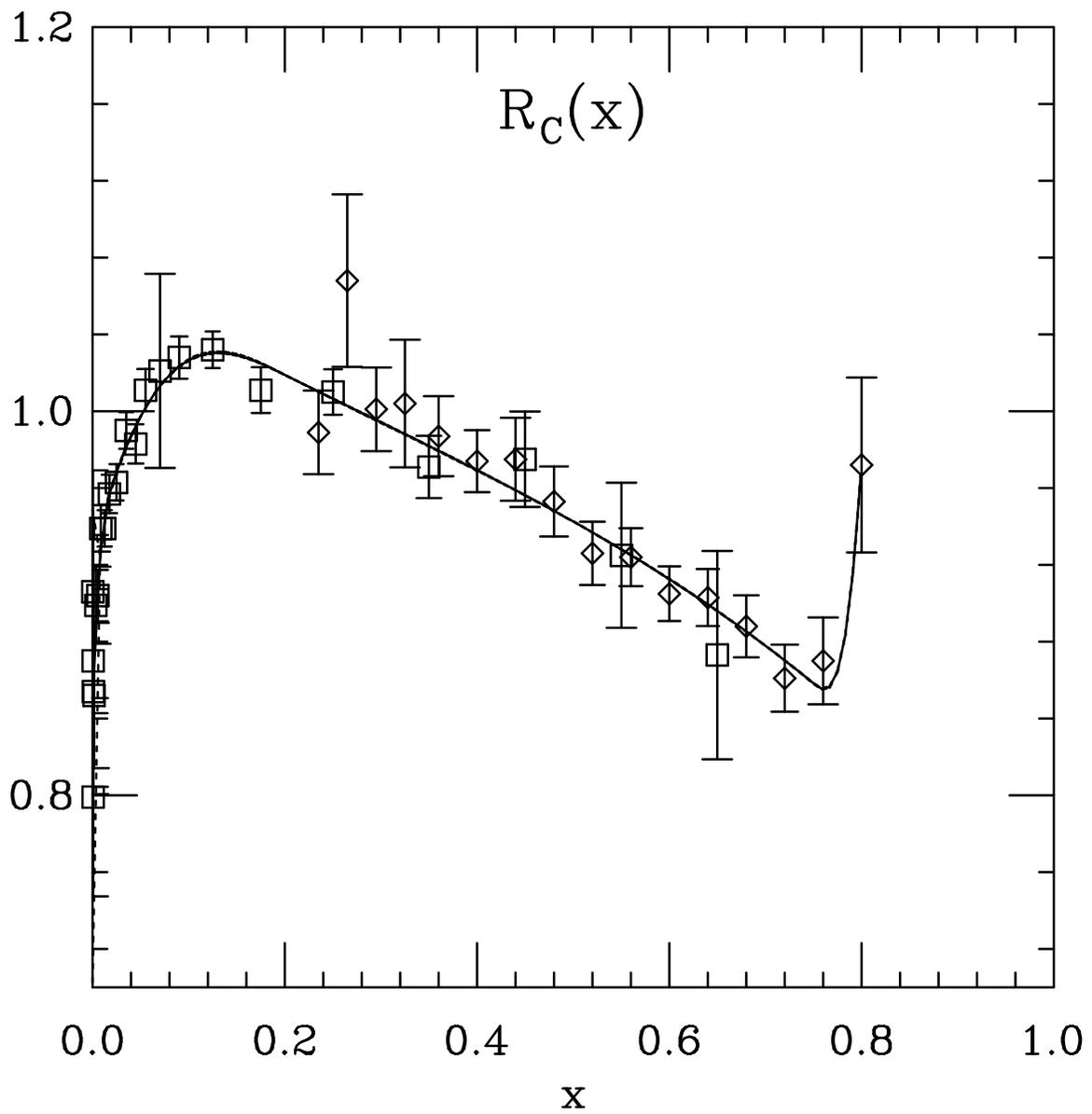
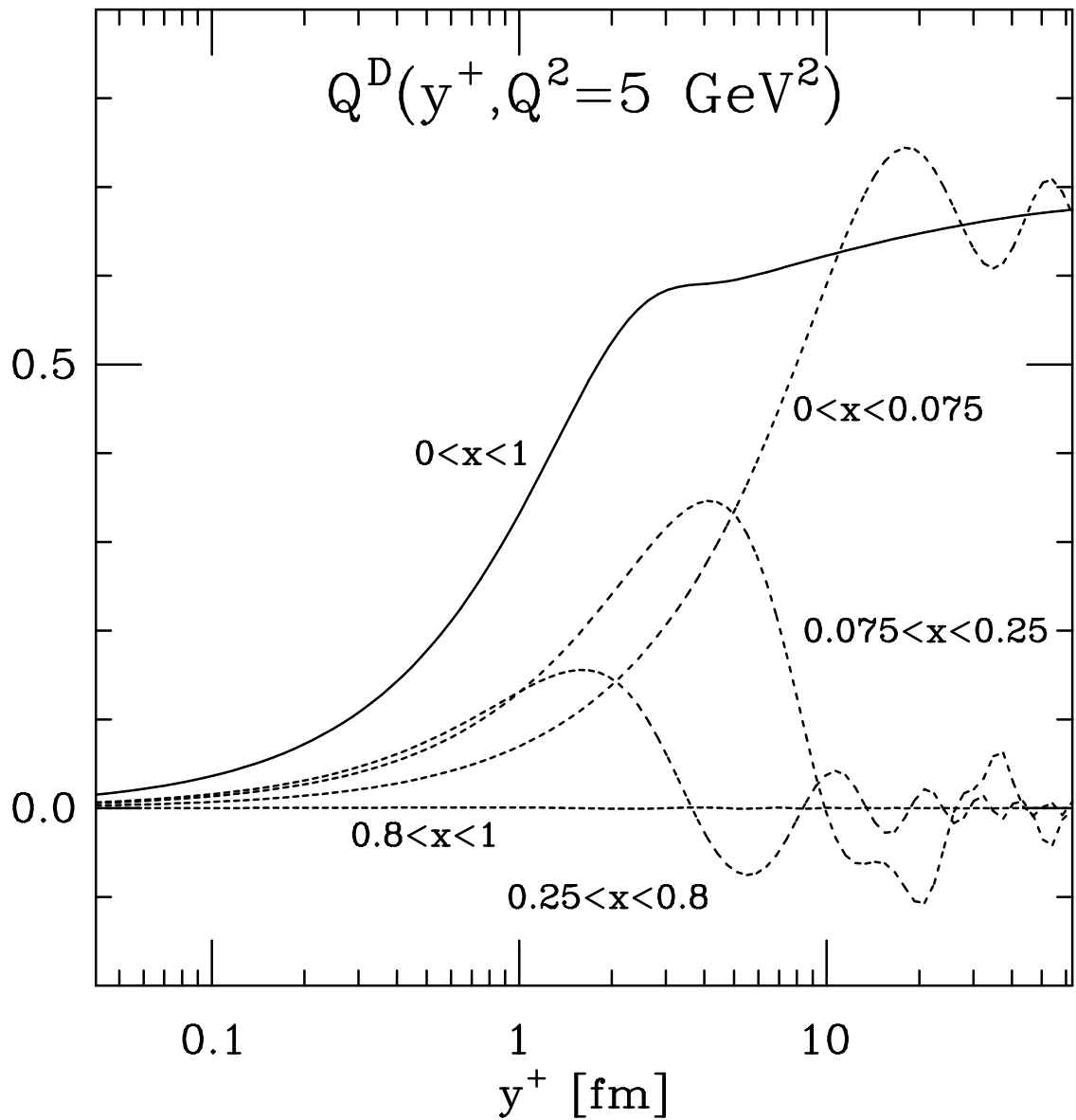
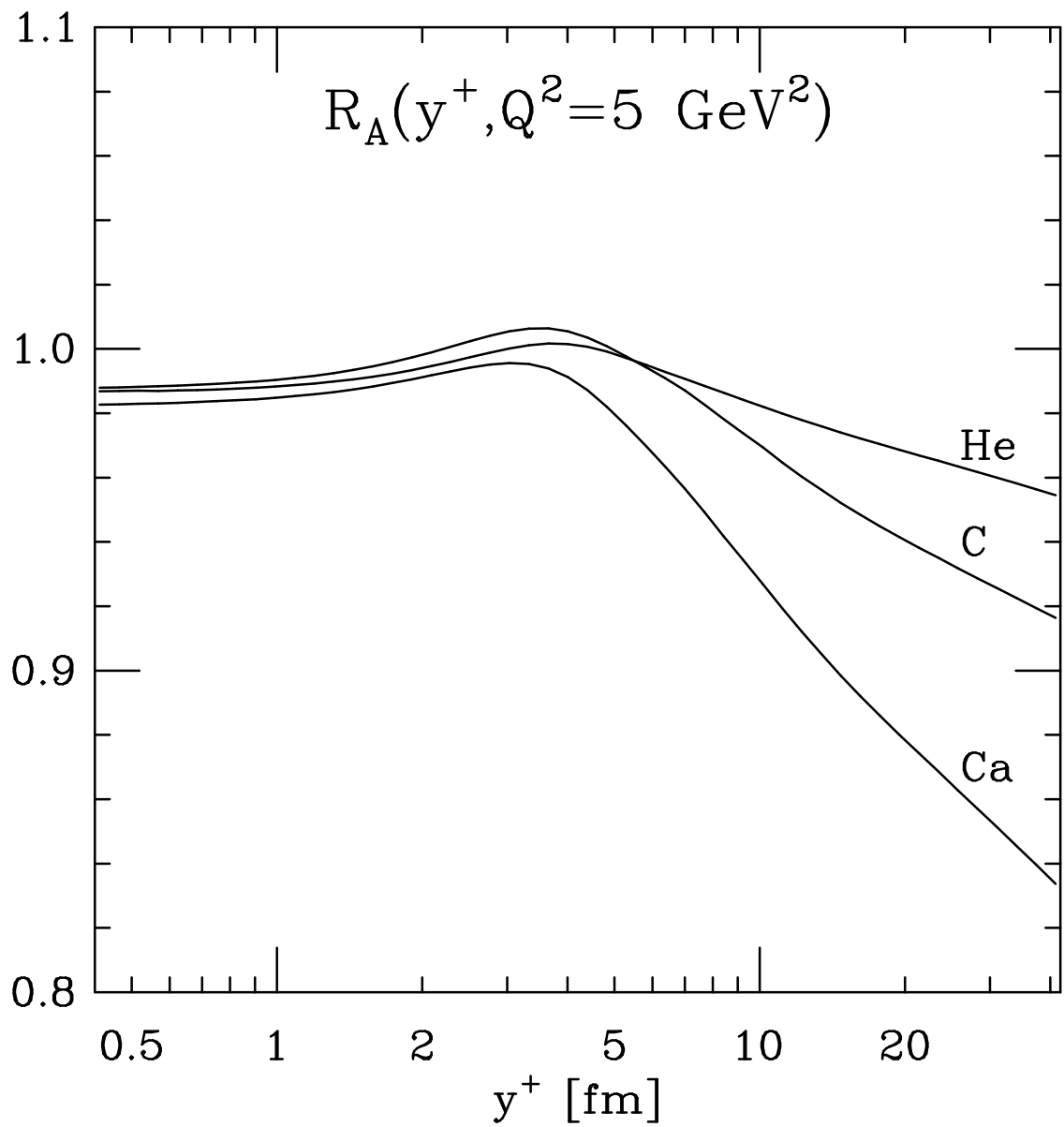
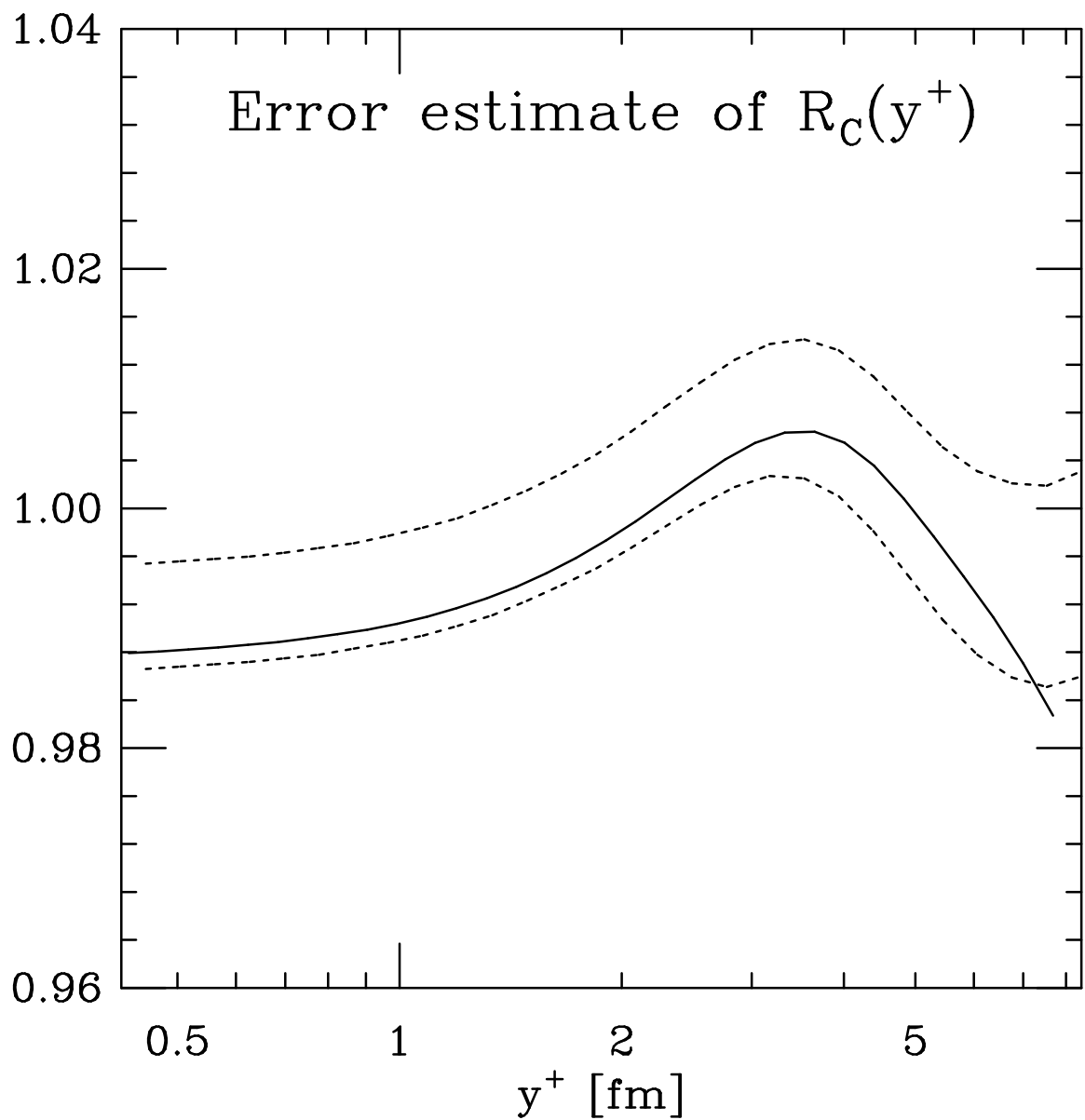
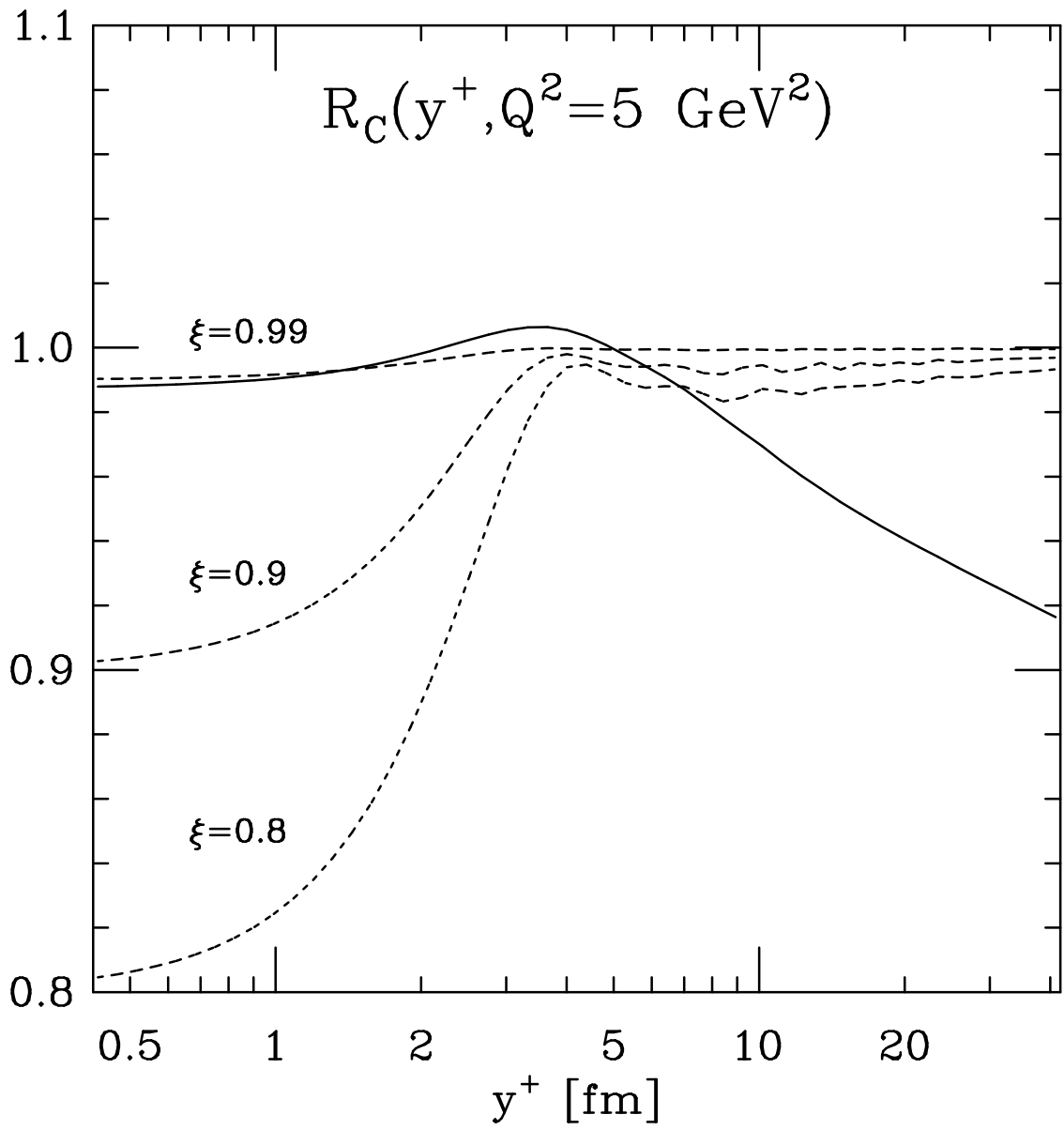
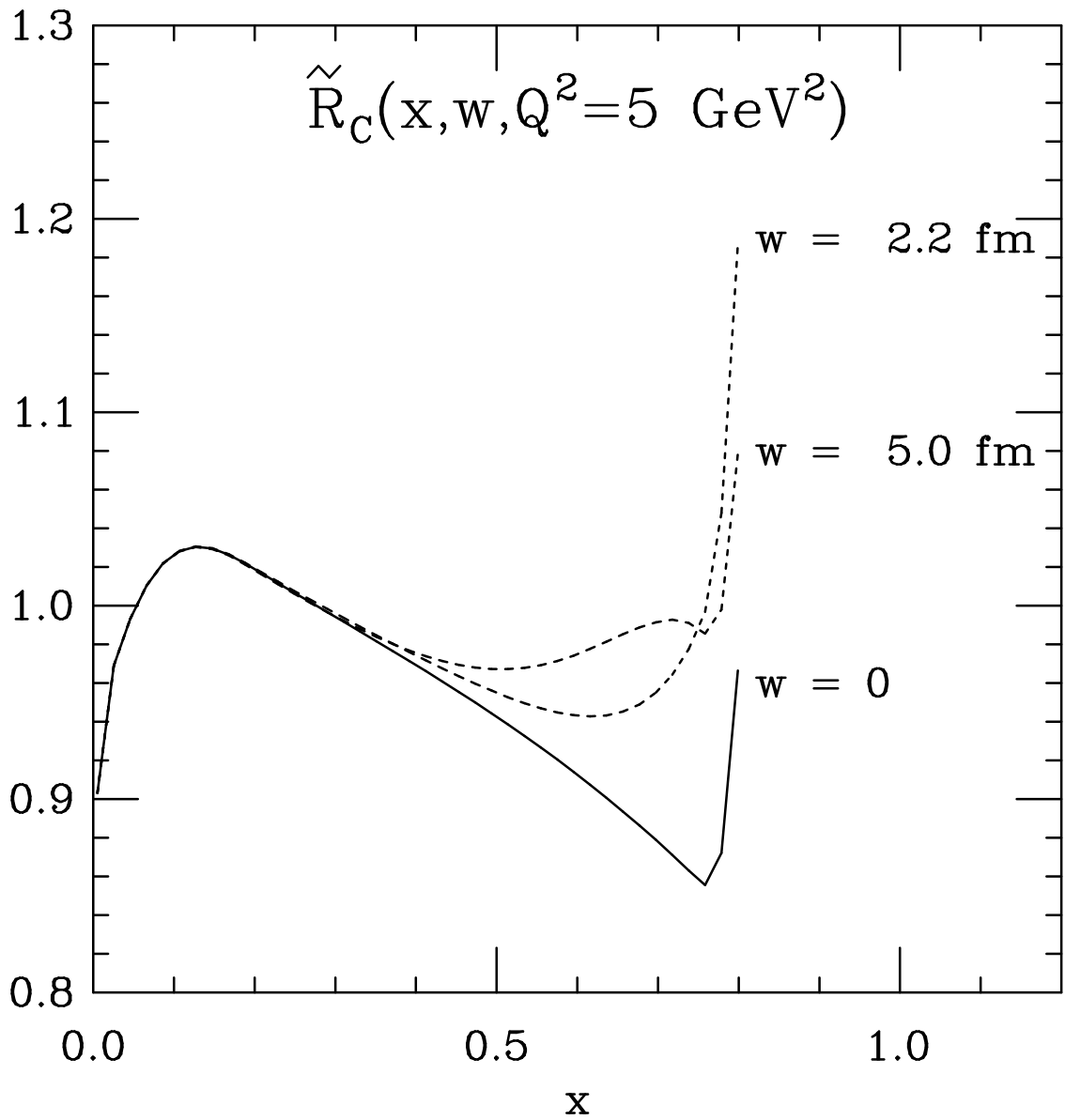
\end{document}